\documentclass[nofootinbib,aps]{revtex4}

\usepackage{amssymb,amstext,amsmath,amsthm}
\usepackage[dvips]{graphicx}
\usepackage{latexsym}
\usepackage{psfrag}
\usepackage{amsfonts}
\usepackage{bbm}

\newcommand{\Ref}[1]{(\ref{#1})}

\def\nn{\nonumber}

\newcommand{\eqa}{\begin{eqnarray}}
\newcommand{\neqa}{\end{eqnarray}}
\newcommand{\equ}{\begin{equation}}
\newcommand{\nequ}{\end{equation}}
\renewcommand{\nn}{\nonumber}

\newcommand{\be}{\begin{equation}}
\newcommand{\ee}{\end{equation}}
\newcommand{\bes}{\begin{eqnarray}}
\newcommand{\ees}{\end{eqnarray}}

\newcommand{\N}{\mathbb{N}}
\newcommand{\Z}{\mathbb{Z}}

\newcommand{\R}{\mathbb{R}}
\DeclareMathOperator{\tr}{tr}

\newcommand{\hh}{{\cal H}}

\def\arr{\rightarrow}

\def\om{\omega}
\def\w{\wedge}
\def\la{\langle}
\def\ra{\rangle}

\newcommand{\1}{$\{10j\}$}

\def\f{\frac}

\def\tf{\tl{f}}
\def\tq{\tl{q}}
\def\tp{\tl{p}}
\def\tx{\tl{x}}
\def\tz{\tl{z}}
\def\bq{{\bf q}}
\def\bp{{\bf p}}

\newcommand{\pp}{\partial}

\usepackage{bbm}

\def\Z{{\mathbbm Z}}

\newcommand{\SU}{\mathrm{SU}}

\def\dag{^\dagger}

\let\eps=\epsilon

\def\cD{{\cal D}}
\def\pp{\partial}

\def\tl{\tilde}

\begin{document}

\title{{\large\bf Notes on the Qubit Phase Space and Discrete Symplectic Structures}}

\author{{\bf Etera R. Livine}\footnote{etera.livine@ens-lyon.fr} }
\affiliation{Laboratoire de Physique, ENS Lyon, CNRS UMR 5672, 46 All\'ee d'Italie, 69364 Lyon,
France}

\begin{abstract}

We start from Wootter's construction of discrete phase spaces and Wigner functions for qubits and more generally for finite dimensional Hilbert spaces. We look at this framework from a non-commutative space perspective and we focus on the Moyal product and the differential calculus on these discrete phase spaces. In particular, the qubit phase space provides the simplest example of a four-point non-commutative phase space. We give an explicit expression of the Moyal bracket as a differential operator. We then compare the quantum dynamics encoded by the Moyal bracket to the classical dynamics 
: we show that the classical Poisson bracket does not satisfy the Jacobi identity thus leaving the Moyal bracket as the only consistent symplectic structure. We finally generalizes our analysis to Hilbert spaces of prime dimensions $d$ and their associated $d\times d$ phase spaces.

\end{abstract}

\maketitle




It is well-known that standard quantum mechanics can be mathematically reformulated in almost classical terms using Wigner functions and the Moyal product. This is achieved through the Weyl transform, which maps phase-space functions to Hilbert-space operators and vice-versa. A by-product is a reformulation of quantum mechanics as a non-commutative geometry with a non-commutative (Moyal) product between functions on the phase space.
This construction was initially introduced for systems evolving in a continuum and thus for infinite dimensional Hilbert spaces such as $L^2(\R^n)$. More recently, Wootters and collaborators introduced a similar construction for finite dimensional Hilbert spaces with a discrete phase space and discrete Wigner quasi-probability distributions \cite{wootters1,wootters2,wootters3} (see also \cite{joe}). For instance, the qubit with its two-state Hilbert space is described in term of a four-point phase space presented as a $2\times 2$ lattice.
In the present notes, we do not introduce a new phase space construction or discuss the use of Wootters' formalism in quantum tomography or related applications.
We are interested in Wootters' framework from a point of view of non-commutative geometry and more specifically we focus on the properties of the discrete Moyal product and the related symplectic structure. Indeed this construction provides very simple examples of non-commutative spaces with finite number of points. These are provided with discrete non-commutative $\star$-products, which reflect the non-commutative product between operators on the Hilbert space.

Starting with the single qubit, we introduce its $2\times 2$ lattice phase space and its discrete Moyal product. We define a differential calculus on his discrete space and provide an explicit formula of the Moyal bracket in term of the discrete differential operators. This shows that this phase space is equipped with a discrete symplectic structure. We further compare this Moyal bracket with the natural Poisson bracket defined in term of the introduced differential calculus. They both have a very similar expression but they lead to substantially different dynamics: the Moyal bracket leads to negative probabilities as expected since it describes quantum dynamics, while the Poisson bracket always keep probabilities positive. However, this Poisson bracket is shown not to satisfy the Jacobi identity (associativity) and thus does not provide a consistent symplectic structure. Somehow, this is not surprising since the qubit is a true quantum system with no classical analog: there is no $\hbar\arr0$ limit in which the Moyal product would be approximated by a classical Poisson bracket on the discrete phase space.

In a second step, we generalize our analysis to Hilbert spaces with higher dimension. Focusing on the case of a prime dimension $d$, we follow Wootters' construction and we define the discrete Weyl map and the Moyal product on the discrete phase space. We give explicit formulas for the Moyal product and brackets. We compare these expressions with the standard Poisson and Moyal brackets on the usual two-dimensional continuous $(q,p)$ phase space (for a one-dimensional system). We show that the continuous Moyal bracket evaluated on certain holonomic observables gives exactly the discrete Moyal bracket: the discrete phase space is thus an exact truncation of the standard phase space where the space of functions over the phase space is restricted to (the Hilbert space generated by) these holonomic observables. This correspondence also shows that the Planck constant $\hbar$ somewhat goes effectively as the inverse of the dimension $d^{-1}$. Finally, we define the discrete differential calculus, discuss the free evolution with a ``$p^2$"-Hamiltonian and identify canonical transformations leaving the symplectic structure invariant. Once again, we conclude that the discrete Moyal bracket defines the only consistent symplectic structure and that there does not seem to be any analog of a classical Poisson bracket expressed in term of the discrete differential calculus and satisfying the Jacobi identity.

\section{A Quick Review of the Qubit Phase Space}

We work with a qubit, i.e living in a spin-$\f12$ representation of $\SU(2)$. The Hilbert space $\hh$ is two-dimensional.  We call $X,Y,Z$ the three Pauli matrices~\footnotemark, satisfying the usual Lie algebra commutator $[X,Y]=2iZ$.  These operators $X,Y,Z$ have eigenvalues $\pm 1$. We denote as usual $|\uparrow\ra, |\downarrow\ra$ the two eigenvectors of $Z$, and we introduce the eigenvectors $|\leftarrow\ra,|\rightarrow\ra$. The projectors on these eigenvectors read:
\bes
&& |\uparrow\ra\la\uparrow|\,=\,\f12(I+Z),\quad  |\rightarrow\ra\la\rightarrow|\,=\,\f12(I+X),
\nonumber\\
&&|\downarrow\ra\la\downarrow|\,=\,\f12(I-Z), \quad |\leftarrow\ra\la\leftarrow|\,=\,\f12(I-X).
\ees
\footnotetext{Our convention for the Pauli matrices are:
$$
X=\left(\begin{array}{cc} 0&1\\1&0 \end{array}\right),\quad
Y=\left(\begin{array}{cc} 0&-i\\i&0 \end{array}\right),\quad
Z=\left(\begin{array}{cc} 1&0\\0&-1 \end{array}\right).
$$
}
We define the following four (Hermitian) observables:
\bes
&&D_{-+}\,=\,\f14(I+X-Y-Z),\quad D_{++}\,=\,\f14(I+X+Y+Z),\nonumber\\
&&D_{--}\,=\,\f14(I-X+Y-Z),\quad D_{+-}\,=\,\f14(I-X-Y+Z),
\ees
or more compactly $D_{\eps\eps'}=(I+\eps'X+\eps\eps'Y+\eps Z)/4$. These $D$-operators form a POVM and define an abstract four point space $\pm\pm$. This defines the discrete phase space for the qubit. This construction has the property that the sums of the $D$-operators along the lines of this space give the projectors on the eigenvectors of $X$ and $Z$:
$$
\begin{array}{l}
D_{--}+D_{-+}=|\downarrow\ra\la\downarrow|,\quad D_{--}+D_{+-}=|\leftarrow\ra\la\leftarrow|,\\
D_{+-}+D_{++}=|\uparrow\ra\la\uparrow|,\quad D_{-+}+D_{++}=|\rightarrow\ra\la\rightarrow|.
\end{array}
$$
We also have $D_{--}+D_{++}=(I+Y)/2$ and $D_{-+}+D_{+-}=(I-Y)/2$ along the diagonal lines. In that
sense, the operator $--$ represent the point $(\downarrow,\leftarrow)$ and so on. Since we can not
diagonalize simultaneously $X$ and $Z$, we can not fully distinguish -separate- these four points.
This translates to the fact that the $D$ operators do not commute with each other, e.g
$[D_{--},D_{-+}]=i(X+Y)/8$.

This four point space is our discrete phase space. The eigenvectors $\downarrow,\uparrow$ define
the ``q" coordinate direction while $\leftarrow,\rightarrow$ define the ``p" momentum direction.

These $D$ operators satisfy the following trace identities:
\be
\forall \alpha,\beta=(\pm\pm),\quad\tr D_\alpha =\f12,\quad\tr D_\alpha
D_\beta=\f12\delta_{\alpha\beta}.
\ee
Resulting, the $D$'s form an orthogonal basis for linear operators on $\hh$ and we can decompose
any operator in this basis. We have~:
\be
F=2\sum_\alpha f_\alpha D_\alpha, \,\textrm{ with }\, f_\alpha=\tr FD_\alpha,
\ee
for all operators $F$ on $\hh$.
The factor 2 is simply the dimension of the Hilbert space $\hh$.
Notice that since $\sum_\alpha D_\alpha=I$, we have $\tr F=\sum_\alpha f_\alpha$ for all functions on the phase space.

We can write density matrices in this $D_\alpha$ basis. A density matrix $\rho$, satisfying
$\rho\dag=\rho,\tr\rho =1$, can be naturally decomposed on the Pauli matrices:
$$
\rho=\f12(I+aX+bY+cZ), \quad a,b,c\in\R.
$$
Since $\det\rho=(1-a^2-b^2-c^2c)/4$, the positivity requirement $\rho\ge 0$ means that
$a^2+b^2+c^2\le 1$. We then define the projections $\rho_\alpha=\tr D_\alpha \rho$~:
\bes
&&\rho_{-+}\,=\,\f14(1+a-b-c),\quad \rho_{++}\,=\,\f14(1+a+b+c),\nonumber\\
&&\rho_{--}\,=\,\f14(I-a+b-c),\quad \rho_{+-}\,=\,\f14(1-a-b+c).
\ees
These are the probabilities that the system is located on each of the four points $(\pm\pm)$ of our discrete phase space. We easily check that these probabilities sum up to 1.

Finally, we can define a notion of ``classical" states $\rho$ for which all our projections are
positive, $\forall \alpha,\, \rho_\alpha\ge0$.

\section{Star Product, Differential Calculus and Symplectic Structure}

We define the multiplication between functions (observables) on the discrete phase space as the
operator multiplication between operators on $\hh$. More precisely, for functions $f_\alpha=\tr
D_\alpha F$ and $g_\alpha=\tr D_\alpha G$, we define:
\be
(f\star g)_\alpha=\f12\tr D_\alpha FG.
\ee
The factor $\f12$ is simply due to the factor 2 entering the map between operators and
functions on the phase space. It ensures that the constant function $f_\alpha=1$ is the unity for the $\star$-product. This product is obviously associative but not commutative and we define our discrete Moyal bracket as its commutator~:
\be
\left(\{f,g\}_\star\right)_\alpha=-\f i{2\hbar}\tr D_\alpha [F,G].
\label{Poissondef}
\ee
The $\hbar$-factor is introduced by hand so as to match the expression of the symplectic
structure in term of the Moyal $\star$-product in standard quantum mechanics~\footnotemark.
\footnotetext{We remind the reader that, in standard quantum mechanics, Moyal bracket only matches the Poisson bracket at first order in $\hbar$.}
This bracket is obviously anti-symmetric and satisfies the Leibniz rule and the Jacobi identity by construction (since the operator commutator does)~\footnotemark.
\footnotetext{For the matrix commutator, the Jacobi identity reads simply:
$$
[F,[G,H]]+[G,[H,F]]+[H,[F,G]]=0.
$$}

We give the simple examples of the operators $X$ and $Z$. The projections of $Z$ on the $D_\alpha$'s are~:
$$
Z_{--}=\tr D_{--}Z=-\f12=Z_{-+},\quad Z_{+-}=Z_{++}=+\f12.
$$
Therefore the operator $2Z$ defines the coordinate $q=\pm 1$ in the phase space. Similarly, we
have $X_{--}=X_{+-}=-1/2$ and $X_{-+}=X_{++}=+1/2$, so that the operator $2X$ defines the momentum coordinate $p=\pm 1$. We can compute their Moyal bracket:
$$
(\{q,p\}_\star)_\alpha\,=\,-\f{2i}{\hbar}\tr D_\alpha [Z,X]=\f{4}{\hbar}\tr D_\alpha Y.
$$
The trace of $D_\alpha Y$ gives $\f12$ at the points $\alpha=(--),(++)$ and $-\f12$ on the points $(-+),(+-)$. Therefore, we obtain the following bracket:
\be
\{q,p\}_\star\,=\,2\f{qp}{\hbar}.
\ee
Let us stress that the Planck constant $\hbar$ enters directly in the Moyal bracket because we have put it by hand in our definition \Ref{Poissondef} above. More interesting, we point out that such a quadratic structure for a Poisson bracket is usually the signature for a quantum group (and it gives directly its $R$-matrix).

We now introduce differential operators. Following a standard strategy in non-commutative geometry, we define the derivation with respect to an operator $\Theta$, or equivalently with respect to the corresponding function $\theta$, as the commutator with $\Theta$~:
\be
(\pp_\Theta f)_\alpha= -i\tr D_\alpha[\Theta,F],
\quad\textrm{or equivalently}\quad
\pp_\Theta f=2\hbar\{\theta,f\}_\star.
\ee
$\pp_\Theta$ obviously satisfies the Leibniz rule $\pp(f\star g)=f\star\pp g+\pp f\star g$ since
the commutator $[\cdot,\cdot]$ does.
For instance, we define the derivative $\pp/\pp p$ along the $p$-direction by considering the
operator $\Theta=Z$. A short calculation yields:
\bes
(\pp_Z f)_{--}=(f_{++}-f_{+-}),\quad (\pp_Z f)_{-+}=-(f_{++}-f_{+-}), \nonumber\\
(\pp_Z f)_{+-}=-(f_{-+}-f_{--}),\quad (\pp_Z f)_{++}=(f_{-+}-f_{--}).
\ees
Similarly, $\pp_X$ defines the derivative along the $q$-direction and the derivation with respect to the $Y$-operator gives variations along the diagonal lines~\footnotemark.
\footnotetext{We compute
$$
(\pp_X f)_{--}=(f_{-+}-f_{++}),\quad (\pp_X f)_{-+}=-(f_{--}-f_{+-}), \quad
(\pp_X f)_{+-}=-(f_{-+}-f_{++}),\quad (\pp_X f)_{++}=(f_{--}-f_{+-}),
$$
$$
(\pp_Y f)_{--}=(f_{-+}-f_{+-}),\quad (\pp_Y f)_{-+}=(f_{++}-f_{--}), \quad
(\pp_Y f)_{+-}=-(f_{++}-f_{--}),\quad (\pp_Y f)_{++}=-(f_{-+}-f_{+-}).
$$}
%
%
Looking at the expression of the derivative $\pp_Z$, we notice that these formulas are not exactly as expected since $(\pp_Z f)_{--}$ computes the finite variation in the $p$-direction at the point $(+-)$ instead of at the point $(--)$. To understand this point shift, we look more closely at the translations defined on our discrete phase space.

For instance, $Z$ generates translations along the $p$-direction~:
$$
ZD_{--}Z=D_{-+},\quad ZD_{-+}Z=D_{--},\quad ZD_{+-}Z=D_{++},\quad ZD_{--}Z=D_{+-},
$$
where we remind that $Z^{-1}=Z$. Thus we can define an operator $T_Z$ acting on a function $f$ as $(T_Zf)_\alpha=\tr D_\alpha ZFZ=\tr ZD_\alpha ZF$. This gives $(T_Zf)_{--}=f_{-+}$ and so on. This allows to define another differential operator computing functional variations in the direction generated by $Z$ naturally as $\delta_Z f = (T_Z f -f)$:
\be
(\delta_Z f)_\alpha =\tr D_\alpha ZFZ -\tr D_\alpha F =\tr D_\alpha [Z,F]Z=-\tr D_\alpha [Z,FZ].
\ee
The relation between the two differential operators $\pp_Z$ and $\delta_Z$ is simple: $\pp_Zf$ is equal to $\delta_Z \tf$, with $\tf$ corresponding to the operator $FZ$ while $f$ is given by $F$. This explains the point shift noticed early which is due to that extra-multiplication by $Z$.
The new operator $\delta_Z$ does not satisfy the same Leibniz rule as the differential operators $\pp_\Theta$, but we obtain a (slightly) deformed Leibniz rule:
$$
\delta_Z(fg)\,=\,
(T_Zf)\star \delta_Zg +\delta_Zf\star g.
$$
In the following, we will denote the difference operators in the $q$ and $p$ directions respectively as $\delta_q\equiv\delta_X$ and $\delta_p\equiv \delta_Z$.

\bigskip

We are now ready to give a full explicit expression for the $\star$-product:
$$
(f\star g)_\alpha=\f12\tr D_\alpha FG = 2 \sum_{\beta,\gamma}f_\beta g_\gamma\tr(D_\alpha D_\beta D_\gamma).
$$
We can compute explicitly the traces of cubic polynomials of the $D$-matrices. For $\alpha=(++)$, we give $16\tr(D_\alpha D_\beta D_\gamma)$ in the following table with $\beta$ labeling the columns and $\gamma$ the lines:
$$
\begin{array}{c|cccc}
& ++ & +- & -+ & -- \\ \hline
++ & 5 & 1 &1 &1 \\
+- & 1 & 1 & (-1-2i) & (-1+2i) \\
-+ & 1 & (-1+2i) & 1 & (-1-2i) \\
-- &1 & (-1-2i) & (-1+2i) & 1
\end{array}
$$
This leads to the following formula:
\bes
(f\star g)_{++} &=&
\f58 f_{++}g_{++} \nn \\
&& +
\f18\left(f_{++}g_{+-}+f_{++}g_{-+}+f_{++}g_{--}+f_{+-}g_{++}+f_{-+}g_{++}+f_{--}g_{++}+f_{+-}g_{+-}+f_{-+}g_{-+}+f_{--}g_{--}\right)
\nonumber \\
&& +\f18\left(-1+2i\right)\left(f_{+-}g_{-+}+f_{-+}g_{--}+f_{--}g_{+-}\right) +
\f18(-1-2i)\left(f_{+-}g_{--}+f_{--}g_{-+}+f_{-+}g_{+-}\right).
\ees
We first notice that even if the original functions $f_\alpha$ and $g_\alpha$ are real, their product has a priori a non-trivial imaginary part.
This product can actually be written as $f_{++}g_{++}+[\textrm{difference terms}]$. The symmetric
terms are real while the anti-symmetric terms are purely imaginary and contribute to the Moyal
bracket. More precisely, we have:
\bes
\left(\{f,g\}_\star\right)_{++}&=&
\f1{2\hbar}(f_{+-}g_{-+}+f_{-+}g_{--}+f_{--}g_{+-}-f_{+-}g_{--}-f_{--}g_{-+}-f_{-+}g_{+-}),\nn\\
&=&\f1{2\hbar}\Big{[}(f_{+-}-f_{--})(g_{-+}-g_{--})-(f_{-+}-f_{--})(g_{+-}-g_{--})\Big{]},\\
&=&\f1{2\hbar}\Big{[}\delta_qf\delta_pg-\delta_pf\delta_qg\Big{]}_{--}.
\ees
We notice that we recover the  $\pp_q\w\pp_p$ structure of the usual Poisson bracket, except for
the point shift $(++)\rightarrow(--)$. In fact, one can show~\footnotemark:
\be
\forall \alpha,\quad
\tr D_\alpha[F,G]\,=\,
i\Big{[}
\tr D_\alpha Y(XFX-F)Y \tr D_\alpha Y(ZGZ-G)Y
-\,(F\leftrightarrow G)\,
\Big{]},
\ee
which translates to the more compact formula in terms of functions on the (discrete) phase
space\footnote{Remember that $T_Y=T_ZT_X$.}:
\be
\{f,g\}_\star\,=\,
\f1{2\hbar}
\,T_Y\big{[}
\delta_qf\delta_pg-\delta_pf\delta_qg
\big{]},
\label{starbracket}
\ee
where the product between $\delta f$ and $\delta g$ is the standard (point) product and not the
$\star$ product. As we will see below, the $T_Y$ translation is very important when considering the dynamics on this phase space~: it seems to be the difference between our Moyal $\star$-bracket for the qubit phase space and its classical counterpart. We could not find such a simple formula in term of the other differential operators $\pp_{p,q}$.

\footnotetext{
We first prove the formula $\tr D_\alpha Z=2\tr D_\alpha X\tr D_\alpha Y$ which holds for all permutations of $X,Y,Z$ and which follows from the simple calculations:
$$
\forall \alpha=(\eps\eps'),\quad
2\tr D_{\eps\eps'} X=\eps',\quad
2\tr D_{\eps\eps'} Y=\eps\eps',\quad
2\tr D_{\eps\eps'} Z=\eps.
$$
Then it is straightforward to show that:
$$
\forall \alpha,\quad
\tr D_\alpha[F,G]\,=\,
i\Big{[}
\tr D_\alpha (ZFZ-YFY) \tr D_\alpha (XGX-YGY)
-\,(F\leftrightarrow G)\,
\Big{]},
$$
by checking it on the operator basis $F,G\in\{I,X,Y,Z\}$. More precisely, this identity is trivial satisfied when $F$ or $G$ are $I$ and when $F=G$. Moreover, since the formula is anti-symmetric, we only need to check explicitly three cases, $(F,G)=(X,Y),(Y,Z),(Z,X)$. Finally this equation is equivalent to the expression given above.
}

\section{Evolution: Quantum vs ``Classical"}

A Hamiltonian evolution $d\rho/dt=-\f i\hbar[H,\rho]$ can be translated in terms of functions on
the phase space and $\star$-product:
\be
\f{d\rho}{dt}=\{H,\rho\}_\star.
\ee
For a single qubit, the only non-trivial (Hermitian) Hamiltonian (up to unitary transformation and multiplication by a constant factor) is
$H=X$. As we showed earlier, this corresponds to the function $p$ on the phase space and thus should simply generate motion along the $q$-axis. Using the
previous calculation of $\pp_X$, we get the following equations of motion:
\bes
&& \f{d\rho_{++}}{dt}=(\rho_{+-}-\rho_{--}), \quad \f{d\rho_{-+}}{dt}=(\rho_{--}-\rho_{+-}),  \nn\\
&&\f{d\rho_{+-}}{dt}=(\rho_{-+}-\rho_{++}), \quad \f{d\rho_{--}}{dt}=(\rho_{++}-\rho_{-+}).
\ees
We check the conservation of probabilities along the classical trajectories:
$$
\f{d(\rho_{++}+\rho_{-+})}{dt}\,=\,0\,=\,\f{d(\rho_{+-}+\rho_{--})}{dt},
$$
where a trajectory is given by a fixed $p$ (the second sign) and the two different signs of $q$.
This translates the fact that  eigenstates of $X$ are stable under the evolution.

We can check that these equations of motion naturally lead to negative probabilities.
Physically, this is a signature of a quantum regime. Mathematically, this is due to the point
shift in the discrete derivative $\pp_X$.

To solve the system, we call $A$ the constant of motion: $\rho_{++}+\rho_{-+}=A$ and  $\rho_{+-}+\rho_{--}=1-A$ (since $\tr\rho =1$ is fixed). Inserting this in the equations, we obtain a second order equation for, say, $\rho_{++}$:
\be
\f{d^2\rho_{++}}{dt^2}=A+1-4\rho_{++},
\ee
which finally get solved by $\rho_{++}=K\cos(2t)+(A+1)/4$ where $K$ is another constant of integration (there could be an oscillation in $\sin$ too). Similarly, we derive:
$$
\rho_{-+}=-K\cos(2t)+\f{3A-1}{4},\quad
\rho_{--}=-K\sin(2t)+\f{A-1}{2},\quad
\rho_{+-}=K\sin(2t)+\f{3-3A}{2}.
$$
We check that these probabilities sum to 1 as wanted. Then, due to the oscillatory behavior, we can easily get negative probabilities. For instance, taking $A=1$ and $K=1/2$, we have $\rho_{++}=1$ and all other probabilities equal to 0 at the initial time $t=0$, but $\rho_{--}$ becomes negative as soon as the system starts to evolve.

\medskip

Now, we can compare this dynamics with the ``classical" evolution defined by the following equations of motion:
\bes
&& \f{d\rho_{++}}{dt}=(\rho_{-+}-\rho_{++}), \quad \f{d\rho_{-+}}{dt}=(\rho_{++}-\rho_{-+}),  \nn\\
&&\f{d\rho_{+-}}{dt}=(\rho_{+-}-\rho_{--}), \quad \f{d\rho_{--}}{dt}=(\rho_{--}-\rho_{+-}). \nn
\ees
Such an evolution would keep probabilities positive all the time (if they are positive to begin with). That's why we call ``classical" this choice of dynamics. The main difference between these classical equation of motion is that the two trajectories $(++\leftrightarrow -+)$  and $(-+\leftrightarrow --)$ are completely decoupled, while the true quantum motion actually makes them interfere with each other.
Resulting, the true evolution leads to an oscillatory behavior (in cosine) while this fictitious ``classical" evolution gives a simple exponential damping (since $d\rho_{++}/dt\,=\,A-2\rho_{++}$ and so on).

Moreover, this evolution can be derived from a ``classical" Poisson bracket, defined from our
Moyal bracket $\{\cdot,\cdot\}_\star$ by dropping the $T_Y$-translation from equation
\Ref{starbracket}:
\be
\{f,g\}_{cl}\,\equiv\,
\f1{2\hbar}
\,\big{[}
\delta_qf\delta_pg-\delta_pf\delta_qg
\big{]},
\ee
or equivalently
$$
(\{f,g\}_{cl})_\alpha\,\equiv\, -\f i{2\hbar}\tr D_\alpha Y[F,G]Y.
$$
Due to these extra $Y$-operators, one can check that this ``classical" Poisson bracket does not
satisfy the Jacobi identity unlike the Poisson bracket defined as the commutator of the
$\star$-product. Therefore, at the discrete level, we do not have a choice: only the Moyal bracket and not the ``classical" Poisson bracket defines a proper  Poisson structure (satisfying the Jacobi identity) and thus a consistent symplectic structure.

\section{Going Beyond the Qubit}

Following Wootters' construction for discrete phase space, we consider a $d$-dimensional Hilbert space $\hh$, with $d\ge 3$ a (odd) prime number (it can be generalized to the power of a prime number by working with finite fields), and we construct a $d\times d$ discrete phase space as in the qubit case by introducing a suitable basis for the $d^2$-dimensional space of (Hermitian) operators on $\hh$. Choosing a basis of $\hh$, with vectors $|a\ra$ labeled by $a\in\Z_d$, we introduce the following two unitary operators:
\be
X\,|a\ra=|a+1\ra,\quad
Z\,|a\ra=\,\om^a\,|a\ra,
\ee
where $\om=\exp(2i\pi/d)$ is a $d$-root of unity. Then we define the $d\times d$ grid of operators:
\be
D(p,q)\equiv\,
\f1d\sum_{a,b}\om^{pa-qb}\om^{\f{ab}{2}}X^aZ^b,
\ee
where the (finite) sum is implicitly taken over $a,b\in\Z_d$. We remind the reader that $2$ is invertible in $\Z_d$ and that the division by $2$ is equivalent to the multiplication by the integer $(d+1)/2$. Let us underline the fact that this definition does not work in the qubit case for $d=2$. Using the commutation relation $Z^bX^a=\om^{ab}X^aZ^b$ and the fact that $\om$ is a root of unity, we easily check that these operators are Hermitian and we can compute their traces:
\be
D(p,q)\dag=D(p,q),\quad
\tr D(p,q) =1,\quad
\tr D(p,q)D(r,s)=d\delta_{p,r}\delta_{q,s}.
\ee
We also check that $\sum_{p,q}D(p,q)=d\,I$. For any operator $F$ on $\hh$, we associate an observable $f$ on the discrete phase space:
\be
f{(p,q)}=\f1d\tr F D(p,q),\qquad
F=\sum_{p,q}f(p,q)D(p,q),\qquad
\tr F= \sum_{p,q}f(p,q).
\ee
The constant function $f=1$ corresponds to the trivial operator $F=d I$ proportional to the identity. The operator $F=Z$ gives $f(p,q)= z(p,q)=\f1d\om^q$ defines the space coordinate while the operator $X\dag=X^{-1}$ with $x(p,q)=\f1d\om^p$  defines the momentum coordinate. The fact that the coordinates are given effectively by $(\om^q,\om^p)$ instead of simply $(p,q)$ hints that q-numbers for q=$\om$ are likely to appear naturally in this framework~\footnotemark.

\footnotetext{
q-numbers are usually introduced when dealing with q-deformed quantum groups. They are defined as:
$$
[n]_q\,\equiv\,\f{1-q^n}{1-q},
$$
which converges to $n$ in the classical limit $q\arr1$.
}

We define the $\star$-product as the operator product:
\be
(f\star g){(p,q)}=\f1{d^2}\tr FG D(p,q).
\ee
This $\star$-product is associative and non-commutative and its unity is the constant function $f=1$. We define the corresponding Moyal bracket:
\be
\{f,g\}_\star(p,q)=-\f{i}{\hbar}(f\star g- g\star f) =-\f{i}{\hbar d^2}\tr D(p,q)[F,G],
\ee
where the Planck constant $\hbar$ is inserted by hand.
We can give a more explicit expression for the $\star$-product referring solely to the functions on space phase and not to the operators:
\be
(f\star g){(p,q)}=\f1{d^2}\sum_{r,s,t,u}f(r,s)g(t,u)\tr D(p,q)D(r,s)D(t,u)\,=\,
\f{1}{d^2}\sum_{(r,s),(t,u)}f(p+r,q+s)g(p+t,q+u) \om^{2(st-ru)},
\label{stard}
\ee
which we compute by expanding the $D$-operators in terms of the basis $X^aZ^b$.
We recognize the $\star$-product for the non-commutative torus with deformation parameter $\theta=d/2\pi$.

This provides to a simple expression for the Moyal bracket:
\be
\{f,g\}_\star(p,q)=\f{2}{\hbar d^2}\sum_{(r,s),(t,u)}f(p+r,q+s)g(p+t,q+u)
\sin\f{4\pi}{d}(st-ru).
\ee
We can apply this formulate to the space and momentum coordinates, $z(p,q)=\f1d\om^q$ and $x(p,q)=\f1d\om^p$ and compute the canonical bracket (which actually corresponds simply to the commutator of $Z$ and $X$):
\be
\{z,x\}_\star=
-\f{2}{\hbar}\,zx\,\sin\f{\pi}{d}.
\label{zxbracket}
\ee
Setting $d=2$ in this formula, we have $\sin\pi/2=1$ and we recover (up to a constant normalization factor) the Moyal bracket of the qubit case given earlier. We can compute the bracket of more generic functions:
\be
\{\om^{\alpha q+\beta p},\om^{\gamma q+\delta p}\}_\star=
-\f{2}{\hbar}\,\om^{\alpha q+\beta p}\om^{\gamma q+\delta p}
\,\sin\f{\pi}{d}(\alpha\delta-\beta\gamma).
\ee
We would like to compare this to the standard Poisson bracket of classical mechanics in the continuum, $\{\bq,\bp\}_{st}=1$. More precisely, we consider holonomy-like observables~\footnotemark  $\exp(i\lambda \bq)$ and $\exp(i\lambda \bp)$. A simple calculation yields:
\be
\{e^{i\lambda \bq},e^{i\lambda \bp}\}_{st}=
\,-\lambda^2\,e^{i\lambda \bq}\,e^{i\lambda \bp},
\ee
which is very similar to the formula above computed in our discrete setting. We can push this comparison further and compute the Moyal bracket of these holonomic observables.
\footnotetext{
For dimensional purposes, since space and momentum coordinates do not have the same physical dimension, it would be better to consider observables $\exp(i\lambda_1 \bq)$ and $\exp(i\lambda_2 \bp)$ with the constants $\lambda_1$ and $\lambda_2$ a priori different and independent. In this case, all the formulas given here are still valid with $\lambda^2=\lambda_1\lambda_2$.
}
%
The standard Moyal product for a one-dimensional system is given as a power series in $\hbar$ by~:
\be
\label{usualMoyal}
f\star_{st} g \,=\,
\sum_n\f1n\left(\f{i\hbar}{2}\right)^n\sum_{k=0}^n(-1)^k\left(\begin{array}{c}n\\k\end{array}\right)
(\pp_\bp^k\pp_\bq^{n-k}f)(\pp_\bp^{n-k}\pp_\bq^kg),
\ee
and the Moyal bracket is given by the commutator:
$$
\{f,g\}_{\star st}=\f1{i\hbar}(f\star_{st} g-g\star_{st} f)
\quad\underset{\hbar\arr0}{\longrightarrow}\quad
\{f,g\}_{st}.
$$
Then the Moyal bracket of the holonomic observables gives:
\be
\{e^{i\lambda \bq},e^{i\lambda \bp}\}_{\star st}=
-\f{2}{\hbar}\,e^{i\lambda \bq}\,e^{i\lambda \bp}
\,\sin\f{\hbar \lambda}{2}(\alpha\delta-\beta\gamma).
\ee
Actually this is exactly the same as our discrete Moyal product if the following relation between the dimension $d$ and the coefficient $\lambda$ is assumed:
\be
\label{dhbar}
\f{\hbar \lambda}{2}=\f{\pi}{d}.
\ee
This is equivalent to $\om=\exp(2i\pi/d)=\exp(i\hbar\lambda)$. This means that the discrete phase space structure is simply a truncation of the standard continuum phase space to specific holonomic observables defined above. Then the discrete Moyal bracket defined through the (finite dimensional) matrix commutator is exactly equal to the continuum Moyal bracket. Let us emphasize then that we have correctly considered the matrix commutator as defining the Moyal bracket and not an approximated discrete Poisson bracket.

\medskip

We have understood the relation between the discrete Moyal bracket and the continuum Moyal bracket, thus providing a representation of our discrete Moyal bracket in term of the continuum differential calculus. Instead, we would like to be able to express the discrete bracket in term of a discrete differential calculus, just like in the qubit case.
To this purpose, we look at the translations. It is very easy to check $X$ generates shifts in $q$ while $Z$ leads to shifts in $p$. More precisely, we have:
\be
X^sZ^r\,D(p,q)\,(X^sZ^r)\dag=D(p+r,q+s).
\ee
Thus we can define finite difference operators as in the qubit case~\footnotemark:
\be
(\delta_p f) (p,q)\,\equiv \f1d\tr D(p,q)(Z\dag FZ-F) =\f1d\tr F(ZD(p,q)Z\dag-D(p,q))=f(p+1,q)-f(p,q),
\ee
and
\be
(\delta_q f) (p,q)\,\equiv \f1d\tr D(p,q)(X\dag FX-F) =\f1d\tr F(XD(p,q)X\dag-D(p,q))=f(p,q+1)-f(p,q).
\ee

\footnotetext{
We could also define differential operators of the other type such as
$(\pp_p f)(p,q)=\f1{d}\tr D(p,q)[Z,F]$ and the same for $\pp_q$ in term of $X$. Their action is not as simple as the other operators $\delta_p$ and $\delta_q$ and seem to involve shifts in $p$ and $q$ by $\f12$, which is harder to interpret on the lattice. Indeed, keeping in mind that $z(p,q)=\f1d\om^q$, we compute
$$
(\pp_p f)(p,q)=\f1{d}\tr D(p,q)[Z,F]=(\om^q\star f-f\star \om^q)\,=\,
-\om^q\left(f(p+\f12,q)-f(p-\f12,q)\right).
$$
Let us insist that $\f12$ must be understood as an element of $\Z_d$ and equal effectively to the integer $(d+1)/2$. Indeed $p+\f12$ does not correspond to a point in the lattice close to the original point $(p,q)$ but to a point on the opposite side of the lattice.
}

We would like to express the discrete Moyal product/bracket in terms of these differential operators and compare it with the standard formula of the continuum limit. First starting from equation \Ref{stard}, replacing $\om=\exp(2\pi/d)$ and expanding the exponential, we find a formula similar to the usual formula \Ref{usualMoyal}~:
$$
(f\star g)(p,q)=\f{2}{\hbar d^2}
\sum_n^\infty \f{1}{n!}\left(\f{4i\pi}{d}\right)^n
\sum_{k=0}^n (-1)^k\left(\begin{array}{c}n\\k\end{array}\right)
\sum_{r,s}r^ks^{n-k}f(p+r,q+s)\,\sum_{t,u}t^{n-k}u^kg(p+t,q+u).
$$
This suggests that $1/d$ plays the role of an effective Planck constant, as already implicit in the formula \Ref{dhbar} above, but does not allow to express the product in term of the differential operators. On the other hand, we can compute explicitly the iterated action of the differential operators:
\be
(\delta_p^k\delta_q^l f)(p,q)\,=\,
\sum_{r=0}^k\sum_{s=0}^l (-1)^{k-r}(-1)^{l-s}
\left(\begin{array}{c}k\\r\end{array}\right)\left(\begin{array}{c}l\\s\end{array}\right)
f(p+r,q+s),
\ee
which is easily inverted for $r,s\ge0$:
\be
f(p+r,q+s)= \sum_{k=0}^r\sum_{l=0}^s
\left(\begin{array}{c}r\\k\end{array}\right)\left(\begin{array}{c}s\\l\end{array}\right)
(\delta_p^k\delta_q^l f)(p,q).
\ee
This allows an expression of the Moyal product in term of the $\delta$-differentials, but we have not been able to simplify further in order to make it look like the standard formula.

\medskip


We now move on to the dynamics on the discrete phase space. Let us on the simplest dynamics: the ``free motion" generated by the Hamiltonian $H=(X+X\dag)/2$ which is the equivalent of the standard $p^2$. More precisely, we can't take $H=X$ since the operator $X$ is unitary but not Hermitian.
The eigenvalues~\footnotemark of $X$ are $\om^p$ while the the eigenvalues of $H$ are $\cos\f{2\pi p}{d}$. Moreover we can compute the Hamiltonian function on the phase space lattice:
\be
h(p,q)=\f1d\,\cos\f{2\pi}{d}p
\quad\underset{d\gg1}{\sim}\,\f1d\,\left(1-\f{(2\pi)^2}{d^2}\,p^2\right).
\ee
Thus, besides the pre-factor $\f1d$ and a constant shift, this choice of Hamiltonian gives the standard $p^2$ dynamics in the large dimension limit $d\arr\infty$.
\footnotetext{
The eigenvectors of $X$ are $\sum_a \om^{-pa}|a\ra$ while the eigenvectors of $H=(X+X\dag)/2$ are $\sum_a \sin(2\pi pa/d)\,|a\ra$.
}
The Hamiltonian evolution of the density matrix $d\rho/dt=-\f i\hbar[H,\rho]$ is easily translated in term of the probabilities $\rho(p,q)$ on the phase space using the Moyal bracket~:
\be
\f{d\rho}{dt}=\,d\{h,\rho\}_\star
\quad\Rightarrow\quad
\f{d\rho(p,q)}{dt}=\f{1}{\hbar}\,\sin\f{2\pi p}{d}\,\left(\rho(p,q+\f12)-\rho(p,q-\f12)\right).
\ee
Let us remind the reader that $\f12$ is considered as an element of $\Z_d$ and equal effectively to the integer $(d+1)/2$. Indeed $q+\f12$ does not correspond to a point in the lattice close to the original point $(p,q)$ but to a point at the opposite side of the lattice. We also point out the pre-factor which depends non-trivially on $p$.

We can check that the probabilities $\sum_q \rho(p,q)$ for fixed $p$ are conserved. These correspond to the classical trajectories -motion along the $q$-axis with fixed momentum $p$- or equivalently to the eigenvalues of the Hamiltonian $H$. This could be seen directly from the definition of the $D$-matrices. Indeed, for fixed $p$, the operator
\be
\cD_p\,\equiv\,\sum_q D(p,q)=\sum_a \om^{pa}X^a
\ee
is the projector on the eigenvalue $\om^{-p}$ of the (unitary) operator $X$. It commutes with $X$ and thus with $H$. Therefore, we have:
$$
\sum_q  \f{d\rho(p,q)}{dt}= -\f i\hbar\tr\cD_p[H,\rho]=\f i\hbar\tr[H,\cD_p]\rho=0.
$$
Finally, one can check that such an evolution leads to negative probabilities (while keeping of course the sum of all probabilities normalized to 1) even if the initial probabilities were all positive.

We could introduce a concept of classical evolution by using a simple discrete symplectic bracket:
$$
\{f,g\}_{cl}(p,q)=\f1{\hbar d^2} [\pp_pf\pp_qg-\pp_qf\pp_pg],
$$
in order to define dynamics which would keep the probabilities positive under time evolution.
This could be considered as a first order approximation of the full discrete Moyal bracket.
However such a bracket does not satisfy the Jacobi identity and thus does not define a proper Poisson structure. At the end of the day, the discrete Moyal bracket $\{\cdot,\cdot\}_\star$ seems to be the only consistent choice for a discrete Poisson structure.

One can find a rigorous discussion of ``classical" vs quantum states in \cite{ernesto1,ernesto2}. ``Classical" states are defined as having a positive probability distribution (for all choices of Wigner functions based on a fixed set of mutually unbiaised basis). Further physical explanations
of why such states can be called classical in term of computational speed-ups can be found in \cite{ernesto1}.
Unitaries that preserves this positivity criteria, thus sending ``classical" states onto ``classical" states, are shown to form a subgroup of the Clifford group. The Clifford group being discrete, any continuous unitary flow would necessarily produce negative probabilities and non-``classical" states. Nevertheless, for a fixed (time-independent) Hamiltonian, if we call $\tau$ the first time at which the evolution is given by a Clifford group element (if it exists), then at all times $n\tau$ with $n\in\N$, the evolution would again produce a Clifford group element. Thus, it could be that the evolution looks classical if we look only at the system in term of a fixed time unit. Nevertheless, this last remark is vague and remains purely speculative.

\medskip

Finally, we look at canonical transformation, i.e changes in the space and momentum coordinates that preserve the symplectic bracket. Up to now, we have been working with $z=\om^q$ and $x=\om^p$ - we have dropped the $\f1d$ factor in front of the coordinates since it does not affect the bracket \Ref{zxbracket}. Let us define the new coordinates, with $\alpha,\beta,\gamma,\delta\in\Z_d$:
$$
\tz=\om^{\alpha q+\beta p},\qquad
\tx=\om^{\gamma q+\delta p},
\quad
\textrm{or equivalently:}
\quad
\left(\begin{array}{c}\tq \\ \tp\end{array}\right)
=
\left(\begin{array}{cc}\alpha&\beta\\ \gamma & \delta\end{array}\right)
\left(\begin{array}{c}q\\ p\end{array}\right).
$$
As we have already computed previously, we have~:
\be
\{\tz,\tx\}_\star=
-\f{2}{\hbar}\,\tz\tx\,\sin\f{\pi}{d}(\alpha\delta-\beta\gamma),
\ee
which gives the same bracket \Ref{zxbracket} as initially as soon as $(\alpha\delta-\beta\gamma)=1$, i.e the determinant of the change of variable $(p,q)\arr(\tp,\tq)$ is 1. This is the behavior expected from classical mechanics: it ensures that the map $(p,q)\arr(\tp,\tq)$ and thus $(z,x)\arr(\tz,\tx)$ is one-to-one. As shown in \cite{wootters3}, such a canonical change of variable actually corresponds to a unitary change of basis up to a phase, i.e there exists a unitary matrix $U$ depending on $\alpha,\beta,\gamma,\delta$ such that:
$$
UD(p,q)U\dag =\,\om^{\f12(\tp\tq-pq)}\,D(\tp,\tq).
$$
This concludes our analysis of the symplectic structure of these discrete phase spaces.

\section*{Conclusion}

We looked at Wootters' construction of discrete phase spaces and discrete Wigner functions for quantum mechanics on finite-dimensional Hilbert spaces. We studied the induced Moyal product and the resulting discrete symplectic structure. Considering these phase spaces as simple examples of non-commutative geometries, we defined a discrete differential calculus and expressed the Moyal bracket in terms of these differential operators. We showed that the Moyal bracket satisfies all the same properties as the usual Moyal product in the continuum. Actually, we further proved that, for odd prime dimensions, the continuous Moyal bracket evaluated on certain holonomic observables gives exactly the discrete Moyal bracket. Resulting, the discrete phase space appears to be an exact truncation of the standard phase space.

We also discussed the dynamics induced by the discrete Moyal bracket. The Hamiltonian evolution naturally leads to negative (pseudo-)probabilities, which is the signature of the quantum regime. This is expected since the Moyal bracket was constructed to describe exactly the quantum evolution on the finite Hilbert space. We show that one could construct using the discrete differential calculus a kind of classical Poisson bracket for which probabilities would stay positive. However, it turns out that such classical Poisson bracket does not satisfy the Jacobi identity (associativity) and thus does not define a consistent symplectic structure. This means that the discrete Moyal bracket remains the only consistent symplectic structure at the discrete level: we are in the deep quantum regime with no equivalent of the $\hbar\arr 0$ limit and no analog of the Poisson bracket.

Finally, we only considered discrete two-dimensional phase spaces. It would be interesting to generalize our analysis to higher dimensional cases with a quantum system evolving in more than one space dimension. In fact, ,we can simply take the tensor product of the discrete phase spaces considered here. This is actually the Wootters' prescription for Hilbert spaces with non-prime dimension: if the Hilbert space dimension $d$ is non-prime, we consider its prime number factorization $d=d_1^{\alpha_1}..d_n^{\alpha_n}$ and we construct the phase space as the tensor product of the discrete phase spaces associated to each prime factor $d_i^{\alpha_i}$. This way, we get a $2n$-dimensional discrete phase space.
Nevertheless, it would be interesting to obtain a construction where the space dimensions are more intertwined with each other such that the space itself is endowed with a non-commutative structure. Maybe, this is given by the applying the same construction for non-prime dimension $d$ than for a prime dimension without using the tensor product of its prime factors.
We would then get a natural working example of quantum mechanics on a finite non-commutative space with a consistent symplectic structure and Moyal product.

\section*{Acknowledgements}

I am grateful to Joseph Emerson for many discussions and explanations on quantum mechanics and Wootters' discrete phase space constructions.


\end{document}